\documentclass[authoryear,preprint,12pt]{elsarticle}
\usepackage{graphicx} 
\usepackage{amssymb}
\usepackage{amsmath}
\usepackage{mathrsfs}
\usepackage{cite}
\usepackage{amsfonts}
\usepackage{latexsym}
\usepackage[dvipsnames,usenames]{color}
\usepackage[colorlinks=true,
pdfstartview=FitV, linkcolor=cyan,
citecolor=cyan, urlcolor=cyan]{hyperref}
\usepackage{booktabs}
\usepackage{multirow}
\usepackage{caption}
\newcommand{\prova}{\par\noindent\textbf{Proof.} }
\newcommand{\fineprova}{\phantom{A}\hfill{$\blacksquare$}\goodbreak\medskip}

\newcommand{\kernel}{\phi}

\newcommand{\TANG}{T}

\newcommand{\conf}{{\BOmega}}

\newcommand{\coppia}[2]{(#1\,,#2)}

\newcommand{\xx}{x}

\newcommand{\Bdelta}{\boldsymbol{\delta}}

\newcommand{\BOmega}{\boldsymbol{\Omega}}

\newcommand{\TT}{T}

\newcommand{\scalar}[2]{{\langle}\kern.1em#1,#2\kern.1em{\rangle}}
\newcommand{\scalarC}[2]{{\langle}\kern.1em#1,#2\kern.1em{\rangle}_\conf}
\newcommand{\scalarT}[2]{{\langle}\kern.1em#1,#2\kern.1em{\rangle}_\TT}
\newcommand{\scalarTC}[2]{{\langle}\kern.1em#1,#2\kern.1em{\rangle}_{\TANG\conf}}

\newcommand{\sub}[1]{{}_{\lower2pt\hbox{$\scriptstyle#1$}}}

\newcommand{\equaldef}{:=}

\numberwithin{theorem}{section}
\numberwithin{proposition}{section}
\numberwithin{lemma}{section}
\numberwithin{remark}{section}
\numberwithin{definition}{section}
\usepackage{lineno}

\usepackage[demo]{adjustbox}
\usepackage{xcolor}
\usepackage{atbegshi}
\AtBeginDocument{\AtBeginShipoutNext{\AtBeginShipoutDiscard}}	
\PassOptionsToPackage{hyphens}{url}\usepackage{hyperref} 

\begin{document}
	\pagestyle{empty} 
	\begin{titlepage}
		\color[rgb]{.4,.4,1}
		\hspace{5mm}

		\bigskip
		
		\hspace{15mm}
		\begin{minipage}{10mm}
			\color[rgb]{.7,.7,1}
			\rule{1pt}{226mm}
		\end{minipage}
		\begin{minipage}{133mm}
			\vspace{10mm}        
			\color{black}
			\sffamily
			\LARGE\bfseries Limit behaviour of Eringen's two-phase elastic beams  \\[-0.3\baselineskip]   \\[-0.3\baselineskip] 
			
			\vspace{5mm}
			{\large {Preprint of the article published in \\[-0.4\baselineskip] European Journal of Mechanics - A/Solids \\[-0.1\baselineskip] 89, August–October 2021, 104315}} 
			
			\vspace{10mm}        
			{\large Marzia Sara Vaccaro,\\[-0.4\baselineskip] \textsc{Francesco Paolo Pinnola}, \\[-0.4\baselineskip]  Francesco Marotti de Sciarra,\\[-0.4\baselineskip] Raffaele Barretta} 
			
			\large
			
			\vspace{40mm}
			\vspace{5mm}
			
			\small
			\url{https://doi.org/10.1016/j.euromechsol.2021.104315}
			
			\textcircled{c} 2021. This manuscript version is made available under the CC-BY-NC-ND 4.0 license \url{http://creativecommons.org/licenses/by-nc-nd/4.0/}
			\hspace{30mm} 
			\color[rgb]{.4,.4,1} 
		\end{minipage}
	\end{titlepage}


\begin{frontmatter}






\title{Limit behaviour of Eringen's two-phase elastic beams}


\author[label1]{Marzia Sara Vaccaro, Francesco Paolo Pinnola, 
Francesco Marotti de Sciarra, Raffaele Barretta}

\address[label1]{
Department of Structures for Engineering and Architecture,\\ 
University of Naples Federico II, 
via Claudio 21, 80125 - Naples, Italy\\
e-mail: rabarret@unina.it}


\begin{abstract}

In this paper, the bending behaviour of small-scale Bernoulli-Euler beams is investigated
by Eringen's two-phase local/nonlocal theory of elasticity.
Bending moments are expressed in terms of elastic curvatures by a convex combination of 
local and nonlocal contributions, that is a combination with non-negative scalar coefficients 
summing to unity. 
The nonlocal contribution is the convolution integral of the elastic curvature field with a suitable averaging kernel characterized by a scale parameter.
The relevant structural problem, well-posed for non-vanishing local phases, is preliminarily formulated and exact elastic solutions of some simple beam problems are recalled.
Limit behaviours of the obtained elastic solutions, analytically evaluated, studied and diagrammed,
do not fulfill equilibrium requirements and kinematic boundary conditions.
Accordingly, unlike alleged claims in literature, such asymptotic fields cannot be assumed as solutions of the purely nonlocal theory of beam elasticity.
This conclusion agrees with the known result which the elastic equilibrium problem of beams 
of engineering interest formulated by Eringen's purely nonlocal theory admits no solution.
\end{abstract}


\end{frontmatter}

\section{Introduction}
\label{sec: intro}

Over the past decades, growing interest in nanotechnological applications has required a deep knowledge of mechanical behavior of micro- and nano-structures conceived as basic components of smaller and smaller electro-mechanical devices. Among the wide variety of micro/nano-electro-mechanical systems there are biosensors \citep{Soukarie2020}, actuators \citep{LuP2019}, resonators \citep{Chorsi2018}, DNA switches \citep{Chao2020}, porous structures \citep{Jankowski2020,Malikan2020b}, valves/pumps \citep{Banejad2020}, advanced and composite structures \citep{Sedighi2020,Zur2020}, energy harvesters \citep{Ghayesh2020}.

Micro/nano-structures exhibit technically significant size effects which therefore have to be taken into account to accurately model and design new-generation small-scale devices  \citep{GhayeshIJESciS2019}. 
To this end, nonlocal continuum mechanics  
\citep{Polizzotto2001,Polizzotto2002,Bazant2002,Borino2003}
has been conveniently exploited by the scientific community to simulate complex scale phenomena in place of atomistic approaches which are computationally expensive.  

One of the first theories of nonlocal elasticity is the strain-driven model introduced
by \citet{Eringen1983} in the wake of seminal contributions by \citet{Rogula1965,Kroner1967,Rogula1982}. 
According to the strain-driven model, the nonlocal
stress field is the convolution integral between elastic strain and an averaging kernel depending on a characteristic length.
Originally exploited by Eringen to deal with screw dislocations and surface waves, when applied to structural problems involving bounded domains, Eringen's strain-driven model
leads to ill-posed structural problems due to incompatibility
between nonlocal and equilibrium requirements \citep{Romano2017}.

An effective strategy to get well-posed problems
consists in formulating a new nonlocal model in which the roles of
stresses and strains are swapped \citep{RomanoIntModels2017}. 
The stress-driven nonlocal approach has been successfully applied to a wide range of problems of nanotechnological interest
\citep[see e.g.][]{BarrettaCanJCOMB2018,Roghani2020,Zhang2020,BarrettaMRC2018,Pinnola2020,AnsariActaMechSin2018,Barretta2019}.

Another possible strategy to overcome ill-posedness of strain-driven model consists in formulating a two-phase local/nonlocal
model, first introduced in \citep{Eringen1972,Eringen1987} and recently applied in \citep{PisanoFuschi2003,Khodabakhshi2015,Wang2016,Eptaimeros2016,BarrettaPhysicaE}.
In particular, the mixture strain-driven model is based on a convex combination of local and nonlocal phases by means of a mixture parameter, where the nonlocal phase is
the Eringen's integral convolution. Based on two parameters, the two-phase theory is able to model a lot of structural problems. 
However, ill-posedness is eliminated only for positive mixture parameters
\citep{RomanoIntModels2017}. 

\break 

Indeed, for vanishing local phases, Eringen's purely nonlocal strain-driven law is recovered, and hence inconsistencies are expected to occur for structural nonlocal solutions of the limiting elastic equilibrium problem.

In this regard, inaccurate conclusions on the matter are still present in current literature
\citep[see e.g.][]{Mikhasev2020,Mikhasev2021} 
by improperly defining limiting cases of free vibration problems as solutions of elastodynamic problems based on purely nonlocal strain-driven continuum model
which admits no solution \citep{Romano2017,FernandezZaera2017,Fathi2017,Vila2017,Zhang2017,ZhuLi2017,Karami2018, BarrettaCanJCOMB2018,PisanoZamm2021}.

Inconsistencies of limiting solutions of structural problems based on Eringen's two-phase theory will be 
definitely provided in the present research.
The plan is the following. 
The two-phase integral model for slender beams is recalled in Sec.\ref{sec: TwoPhaseT} and the equivalent differential problem is also provided. 
The purely nonlocal strain-driven law is then derived and discussed in Sec.\ref{sec: LimitingCase}. 
Exact structural solutions based on two-phase theory are given in Sec.\ref{sec: CaseStudies} for simple applicative cases. 
Analytical limiting solutions, presented in Sec.\ref{sec: Inconsistencies}, are shown to be in contrast with equilibrium and kinematic boundary conditions. 
Consequently, such fields cannot be assumed as solutions of structural problems formulated according to Eringen's purely nonlocal strain-driven model which is thus confirmed as inapplicable to nanocontinua of technical interest.

\section{Two-phase local/nonlocal theory for elastic beams}
\label{sec: TwoPhaseT}

\def\s{s}
\def\f{f}
\def\m{\alpha}
\def\lc{c}
\def\kernel{\phi}
\def\op{\mathcal{L}_x}
\def\opBa{\mathcal{B}_0}
\def\opBb{\mathcal{B}_L}
\def\arg{\big{(}\f - \m\: \s \big{)}}
\def\disp{v}

Let us consider a slender straight beam under flexure of length $\, L \,$. 
Beam and bending axes are denoted by $\,x\,$ and $\,y\,$ respectively. 
According to Bernoulli-Euler theory, the total curvature field $\, \chi: [0, L] \mapsto \Re \,$
associated with the transverse displacement field $\,\disp: [0, L] \mapsto \Re\, $ 
is expressed by%
\footnote{
The symbol $\,\partial_\xx^n\,$ denotes $\,n$-times differentiation along the beam axis $\,\xx\,$.
}
\begin{equation}
\chi = \partial_x^2 \disp
= \chi^{el}+\chi^{nel} 
\label{kin}
\end{equation}
with $\, \chi^{el} \,$ elastic curvature and $\, \chi^{nel} \,$ all other non-elastic curvature fields. 
Stress fields are described by bending moments $\, M: [0, L] \mapsto \Re \,$ which have to fulfill the differential equation of equilibrium
\begin{equation}
\partial_x^2 M = q
\label{equi}
\end{equation}
with $\, q: [0, L] \mapsto \Re \, $ transversely distributed loading.

According to the two-phase elasticity model \citep{Eringen1972},
the bending moment $\, M\,$ is convex combination of the source field
$\, s\equaldef K \chi^{el}\,$ and of the convolution 
between the source $\, s \,$ and a suitable averaging kernel $\, \kernel_{\lc} \,$
\begin{equation}
M(x) = \m\: \s(x) + (1 - \m) \: \int_{0}^{L}{\kernel_{\lc}(x, \xi) \: \s(\xi) \:d\xi}    
\label{MixConv}
\end{equation}
where $\, K \,$ is the local elastic bending stiffness, i.e. second moment of the field of Euler-Young elastic moduli $\,E\,$ on beam cross section.

The relation in Eq.\eqref{MixConv} is a Fredholm integral equation of the second kind \citep{Tricomi1985,Polyanin2008}
in the unknown source field $\, \s \,$, with $\,0 \leq \m \leq 1\,$ mixture parameter and $\lc>0$ nonlocal scale parameter. 
The purely nonlocal law  $\displaystyle\, M(x)= \int_{0}^{L}{\kernel_{\lc}(x, \xi) \: K \chi^{el}(\xi) \:d\xi} \,$
is got by setting $\, \m = 0\,$,  while 
for $\, \m = 1\,$ the purely local relation $\, M= K \chi^{el} \,$ is recovered.

The averaging kernel is assumed to be the bi-exponential function 
\begin{equation}
\kernel_{\lc}(x) = \frac{1}{2\lc} \exp\biggl(-\frac{|x|}{\lc}\biggr)\,
\label{kern}
\end{equation}
fulfilling symmetry, positivity and limit impulsivity \citep{Eringen1983}.

\def\s{K \chi^{el}}
\def\f{M}

An equivalent differential formulation \citep{Romano2017} of the two-phase model Eq.\eqref{MixConv}
can be got by observing that the special kernel in Eq.\eqref{kern} is the Green's function of the linear differential operator 
$\,\op\,$ defined by 
\begin{equation}
\op \equaldef 1 - \lc^2\,\partial_x^2
\label{op}
\end{equation}
Now, rewriting Eq.\eqref{MixConv} as follows
\begin{equation}
\big{(}\f - \m\: \s\big{)}(x) = (1 - \m) \: \int_{0}^{L}{\kernel_{\lc}(x, \xi) \: \s(\xi) \:d \xi}    
\label{MixConv_r}
\end{equation}
and applying the differential operator $\, \op \,$ to Eq.\eqref{MixConv_r}, we get
\begin{equation}
\op\big{(}\f - \m\: \s\big{)}(x) = (1 - \m) \: \s(x)
\label{Diff}
\end{equation}
which is the differential equation equivalent to the two-phase model in Eq.\eqref{MixConv}.

\break

Indeed, denoting by $\,\Bdelta\,$ the Dirac unit impulse,
since $\, \op\kernel_{\lc}(x, \xi) \, = \Bdelta(x, \xi)\,$, we may write
\begin{equation}
\op\big{(}\f - \m\: \s\big{)}(x)=(1 - \m)\int_{0}^{L}{\op\kernel_{\lc}(x, \xi) \: \s(\xi) \:d \xi}=(1 - \m)\s(x)
\end{equation}
As proven in \citep{Romano2017}, the special kernel in Eq.\eqref{kern} satisfies the following
homogeneous boundary conditions 
\begin{equation}
\setlength{\jot}{8pt}
\left\{
\begin{aligned}
&\opBa\kernel_{\lc}\,|_0 = 0   \\
&\opBb\kernel_{\lc}\,|_L = 0
\end{aligned}
\right.
\label{BCk}
\end{equation}
where $\, \opBa \equaldef 1 - \lc \, \partial_x \,$ and $\, \opBb \equaldef 1 + \lc \, \partial_x \,$ are differential operators defined at the boundary.
By applying $\, \opBa \,$ and $\, \opBb \,$ to Eq.\eqref{MixConv_r} we get 
\begin{equation}
\setlength{\jot}{8pt}
\left\{
\begin{aligned}
&\opBa\arg\,|_0 = 0 \\
&\opBb\arg\,|_L = 0
\end{aligned}
\right.
\label{BCSeq}
\end{equation}
which are the constitutive boundary conditions associated with Eq.\eqref{Diff}.

Finally, the equivalent differential problem in Eqs.\eqref{Diff}-\eqref{BCSeq} can be 
explicitly formulated as follows
\def\s{(K \chi^{el})}
\begin{equation}
\frac{\f (x)}{\lc^2} -\,\partial_x^2\f(x)
=\frac{\s (x)}{\lc^2} - \m\:\partial_x^2\s(x)
\label{EqDiff}
\end{equation}
\begin{equation}
\setlength{\jot}{2pt}
\left\{
\begin{aligned}
&\partial_x \f (0) - \frac{1}{\lc}\,\f (0) = \m\: \bigg{(} \partial_x \s(0) - \frac{\s(0)}{\lc}\bigg{)} \,
\\
&\partial_x \f(L) + \frac{1}{\lc}\,\f(L) = \m\: \bigg{(} \partial_x \s(L) + \frac{\s(L)}{\lc}\bigg{)}      
\end{aligned}
\right.
\label{BCS}
\end{equation}
%
%
%

\break

\section{Limiting case: purely nonlocal elasticity}
\label{sec: LimitingCase}

For vanishing mixture parameter $\,\m=0\,$, the purely nonlocal elasticity model is recovered, that is
\begin{equation}
\f(x) = \int_{0}^{L}{\kernel_{\lc}(x, \xi) \: \s(\xi) \:d \xi}    
\label{Conv}
\end{equation}
which is the Eringen's purely nonlocal strain-driven law.

\medskip
\noindent \textbf{Remark}. \textit{For any value of $\, \lc  >0 \,$, the integral convolution Eq.\eqref{Conv} admits a unique solution (or no solution) if and only if the following constitutive boundary conditions are satisfied by the bending field \citep{Romano2017}}
\textit{
\begin{equation}
\setlength{\jot}{8pt}
\left\{
\begin{aligned}
&\partial_x \f (0) - \frac{1}{\lc}\,\f (0) = 0
\\
&\partial_x \f(L) + \frac{1}{\lc}\,\f(L) = 0 
\end{aligned}
\right.
\label{BC2}
\end{equation}
}
\textit{Fulfillment of Eq.\eqref{BC2} ensures that differential equation (Eq.\eqref{EqDiff2}) provides the unique solution of the nonlocal problem \citep{Romano2017}}
\textit{
\begin{equation}
\frac{\f (x)}{\lc^2} -\,\partial_x^2\f(x)
=\frac{\s (x)}{\lc^2} 
\label{EqDiff2}
\end{equation}
}
As proven in \citep{Romano2017}, constitutive boundary conditions Eqs.\eqref{BC2} are incompatible with
equilibrium requirements, since they relate bending and shearing fields through the characteristic nonlocal length $\, \lc \,$.
The two-phase model eliminates ill-posedness for $\, \m > 0 \,$. Indeed, left-hand sides of Eq.\eqref{BCS}
are known expressions (generally polynomial), unless of $\,n\,$ integration constants, with $\,n\,$ redundancy degree. Hence, the laws in Eq.\eqref{BCS} are non-homogeneous boundary conditions imposed on the elastic curvature and its first derivative and no incompatibility arises with kinematic boundary constraints since they act on displacement fields and its first derivatives. 
However, as $\, \m \to 0 \,$ the two-phase strain-driven law will tend to an ill-posed
theory and then inconsistencies of limiting solutions are expected to occur. Predicted singularities for $\, \m \to 0 \,$ will be investigated in detail in the sequel.

\break

\section{Case-studies and exact structural solutions}
\label{sec: CaseStudies}
\def\Au{\psi_1}
\def\Ad{\psi_2}
\def\At{\psi_3}
\def\A{\psi}
\def\coppia{\mathcal{M}}
\def\s{\chi^{el}}

This section provides exact closed-form solutions of the elastic equilibrium problem of inflected beams 
using Eringen's two-phase model Eqs.\eqref{EqDiff}-\eqref{BCS}.

The solution procedure is summarized as follows.
\begin{itemize}

\item
\textbf{Step 1.}
Solution of the differential equilibrium condition in Eq.\eqref{equi} equipped with standard natural boundary conditions to get the bending moment as functions of $n$ integration constants, with $n$ standing for redundancy degree.
For statically determinate beams, the equilibrated bending field is univocally determined by equilibrium requirements.

\item
\textbf{Step 2.}
Inserting the obtained bending field $\, M \,$ in the equivalent differential problem governed by Eqs.\eqref{EqDiff}-\eqref{BCS} to get the nonlocal
elastic curvature field $\, \chi^{el} \,$.

\item
\textbf{Step 3.}
Detection of nonlocal displacement field $\disp$ by solving Eq.\eqref{kin} with prescription of $\, 3 + n \,$ standard essential boundary conditions.

 \end{itemize}

The solution procedure above is applied to solve the nonlocal elastostatic problems below.
The non-elastic curvature $\,\chi^{nel}\,$ is assumed to vanish, so that, 
by virtue Eq.\eqref{kin}, total and elastic curvatures are coincident 
$\,\chi=\chi^{el}\,$.

\medskip
\textit{Cantilever under concentrated couple at free end}. 

Let us consider a cantilever of length $\, L \,$ and uniform bending stiffness  $\, K \,$ under a concentrated couple $\, \coppia \,$ applied at free end. 

The bending moment field $\, M \,$ is obtained by the differential equilibrium equation $\, \partial_x^2M = 0 \,$ equipped with standard natural boundary conditions $\, M(L) = \coppia \,$ and $\, \partial_xM(L) = 0 \,$, so that Eq.\eqref{EqDiff} becomes
\begin{equation}
\frac{\coppia}{\lc^2}
=\frac{K \s (x)}{\lc^2} - \m\:K \partial_x^2\s(x)
\label{Diff1}
\end{equation}
supplemented with the constitutive boundary conditions
\begin{equation}
\setlength{\jot}{8pt}
\left\{
\begin{aligned}
&- \coppia\, = \m\:K \bigg{(}\lc \, \partial_x \s(0) - \s(0)\bigg{)} \,
\\
&\coppia\, = \m\: K\bigg{(}\lc \, \partial_x \s(L) + \s(L)\bigg{)}      
\end{aligned}
\right.
\label{BCS1}
\end{equation}
Solving the differential problem in Eqs.\eqref{Diff1}-\eqref{BCS1} provides the elastic curvature $\, \s \,$. Also, Eq.\eqref{kin} with prescription of standard essential boundary conditions $\, \disp(0) = 0\,$ and $\, \partial_x\disp(0) = 0\,$, gives the nonlocal displacement field
\begin{equation}
\begin{split}
\disp(x) = &\,\frac{\coppia x^2}{2K}-\frac{\coppia}{\Au K}\, \lc \,(\m-1) \,e^{-\frac{x}{\lc \sqrt{\m}}} \big{(}(x-\lc \sqrt{\m})\, e^{\frac{L+x}{\lc \sqrt{\m}}}+\,\lc\, \sqrt{\m} \,e^{\frac{L}{\lc \sqrt{\m}}}\\
&-e^{\frac{x}{\lc \sqrt{\m}}} (\lc\, \sqrt{\m}+x)+\lc\, \sqrt{\m} e^{\frac{2 x}{\lc\sqrt{\m}}}\big{)}
\end{split}
\label{v1}
\end{equation}
with 
\begin{equation}
\begin{split}
\Au = (\sqrt{\m}+1) \,e^{\frac{L}{\lc \sqrt{\m}}}+\sqrt{\m}-1
\end{split}
\label{psi}
\end{equation}

\medskip
\textit{Cantilever under concentrated force at free end}. 

Let us consider a cantilever of length $\, L \,$ and uniform bending stiffness  $\, K \,$ under a concentrated force $\, F \,$ applied at free end. 

The equilibrated bending moment field $\, M \,$ is defined by $\, \partial_x^2M = 0 \,$ with prescription of standard natural boundary conditions $\, M(L) = 0 \,$ and $\, \partial_xM(L) = - F \,$, so that Eq.\eqref{EqDiff} becomes
\begin{equation}
\frac{F (L - x)}{\lc^2}
=\frac{K \s (x)}{\lc^2} - \m\:K \partial_x^2\s(x)
\label{Diff2}
\end{equation}
equipped with the constitutive boundary conditions
\begin{equation}
\setlength{\jot}{8pt}
\left\{
\begin{aligned}
&- F \lc - F L\, = \m\:K \bigg{(}\lc \, \partial_x \s(0) - \s(0)\bigg{)} \,
\\
&- F \lc\, = \m\: K\bigg{(}\lc \, \partial_x \s(L) + \s(L)\bigg{)}      
\end{aligned}
\right.
\label{BCS2}
\end{equation}
The unknown elastic curvature $\, \s \,$ is obtained by solving the differential problem in Eqs.\eqref{Diff2}-\eqref{BCS2}. Then, Eq.\eqref{kin} with prescription of standard essential boundary conditions  provides the following nonlocal displacement field
\begin{equation}
\begin{split}
\disp(x) =\, &\frac{F x^2 \,(3 L-x)}{6 K} + \frac{(\alpha -1) \,c\, F}{\A K} \,e^{-\frac{x}{\sqrt{\alpha } c}}\,(\sqrt{\alpha } \,\Au\, c^2 \,(e^{\frac{x}{\sqrt{\alpha } c}}-1)\, (e^{\frac{L}{\sqrt{\alpha } c}}+e^{\frac{x}{\sqrt{\alpha } c}}) \\
&-\Ad \,L \,x\, e^{\frac{x}{\sqrt{\alpha } c}}+\,\At\, c)
\end{split}
\label{v2}
\end{equation}
with
\begin{equation}
\begin{split}
&\A = (\sqrt{\alpha }+1)^2\, e^{\frac{2 L}{\sqrt{\alpha } c}}-(\sqrt{\alpha }-1)^2 \\
&\Ad = \sqrt{\alpha }+(\sqrt{\alpha }+1)\, e^{\frac{2 L}{\sqrt{\alpha } c}}-1  \\
&\At = -(\sqrt{\alpha }+1) \,\sqrt{\alpha }\, L\, e^{\frac{2 L}{\sqrt{\alpha } c}}+(\alpha -\sqrt{\alpha }) \,L\, e^{\frac{2 x}{\sqrt{\alpha } c}}-2 \sqrt{\alpha }\,x\, e^{\frac{L+x}{\sqrt{\alpha } c}}+\\
&(\sqrt{\alpha }+1) (\sqrt{\alpha } L-x) \,e^{\frac{2 L+x}{\sqrt{\alpha } c}}-(\sqrt{\alpha }-1) \,e^{\frac{x}{\sqrt{\alpha } c}} (\sqrt{\alpha } L+x)
\end{split}
\label{psi}
\end{equation}

\medskip
\textit{Simply supported beam under uniformly distributed loading}. 

Let us consider a simply supported beam of length $\, L \,$ and uniform bending stiffness  $\, K \,$ under uniformly distributed loading $\, q \,$.

Bending moment field $\, M \,$ is determined by differential equilibrium equation $\, \partial_x^2M = q \,$ equipped with natural boundary conditions $\, M(0) = 0\,$ and $\, M(L) = 0 \,$, so that Eq.\eqref{EqDiff} becomes
\begin{equation}
\frac{q x (x - L)}{2\lc^2} -\,q
=\frac{K \chi^{el}}{\lc^2} - \m\:K \partial_x^2 \chi^{el}
\label{Diff3}
\end{equation}
supplemented with constitutive boundary conditions
\begin{equation}
\setlength{\jot}{8pt}
\left\{
\begin{aligned}
&- \lc \, \frac{qL}{2}\, = \m \, K \: \bigg{(}\lc \, \partial_x \chi^{el}(0) - \chi^{el}(0)\bigg{)} \,
\\
&\lc \, \frac{qL}{2}\, = \m \, K \:\bigg{(} \lc \,\partial_x \chi^{el}(L) + \chi^{el}(L)\bigg{)}      
\end{aligned}
\right.
\label{BCS3}
\end{equation}
 The elastic curvature $\, \s \,$ is obtained from Eqs.\eqref{Diff3}-\eqref{BCS3}; then, Eq.\eqref{kin} together with essential boundary conditions $\, \disp(0) = 0\,$ and $\, \disp(L) = 0\,$, provides the 
 nonlocal displacement field
\begin{equation}
\begin{split}
\disp(x) =\, &\frac{q x\, (L^3-2 L x^2+x^3)}{24 K}-\frac{(\alpha -1) \,c^2 \,q}{2 \Au K} \,e^{-\frac{x}{\sqrt{\alpha }\, c}}\, (-e^{\frac{x}{\sqrt{\alpha }\, c}} \,(2 \,\alpha ^{3/2}\, c^2+L\, (\sqrt{\alpha } c\\
&-\sqrt{\alpha } \,x+x)+(\sqrt{\alpha }-1) \,x^2)+\,e^{\frac{L+x}{\sqrt{\alpha } c}} \,(-2 \,\alpha ^{3/2} \,c^2+L\, (-\sqrt{\alpha }\, c+\sqrt{\alpha }\, x\\
&+x)-(\sqrt{\alpha }+1) \,x^2)+\sqrt{\alpha }\, c\, e^{\frac{L}{\sqrt{\alpha } c}}\, (2 \,\alpha\,  c+L)+\sqrt{\alpha }\, c \,(2\, \alpha\,  c+L) \,e^{\frac{2 x}{\sqrt{\alpha }\, c}})
\end{split}
\label{v3}
\end{equation}

\break

\medskip
\textit{Doubly clamped beam under uniformly distributed loading}. 

Let us consider a doubly clamped beam of length $\, L \,$ and uniform bending stiffness  $\, K \,$ under uniformly distributed loading $\, q \,$.

The bending moment field $\, M \,$ obtained by the differential equilibrium equation $\, \partial_x^2M = q \,$ is a function of two integration constants, $\, a_1\,$ and $\, a_2\,$. Hence, Eq.\eqref{EqDiff} in terms of displacement field $\,\disp\,$  becomes
\begin{equation}
\frac{q x^2}{2} + a_1 x + a_2 - \lc^2 q 
= K \partial_x^2\disp - \m\:\lc^2 \, K \partial_x^4 \disp
\label{Diff4}
\end{equation}
supplemented with constitutive and kinematic boundary conditions
\begin{equation}
\setlength{\jot}{8pt}
\left\{
\begin{aligned}
&\lc \,a_1 - a_2 = \m K \,(\lc \, \partial_x^3 \disp(0) - \partial_x^2 \disp(0)) \\
&\lc\,(qL + a_1) + \frac{qL^2}{2} + a_1L +a_2 = \m K\,(\lc \, \partial_x^3 \disp(L) +\partial_x^2 \disp(L)) \\
& \disp(0) = 0 \\
&\partial_x\disp(0) = 0  \\ 
& \disp(L) = 0 \\
&\partial_x\disp(L) = 0  
\end{aligned}
\right.
\label{BCS4}
\end{equation}
By solving the differential problem in Eqs.\eqref{Diff4}-\eqref{BCS4} the nonlocal displacement field is given by
\def\Aq{\psi_4}
\begin{equation}
\begin{split}
\disp(x) =\, &\frac{q\, x^2 \,(L-x)^2}{24 K} + \frac{c\,(\alpha -1)\, q}{12\, \Aq\, K} (12\, c^2+6 c \,L+L^2) \,e^{-\frac{x}{\sqrt{\alpha } c}} (\sqrt{\alpha }\, c\, L \,e^{\frac{L}{\sqrt{\alpha } c}}\\
&-e^{\frac{L+x}{\sqrt{\alpha } c}} (\sqrt{\alpha }\, c L-L x+x^2)+e^{\frac{x}{\sqrt{\alpha } c}} (-\sqrt{\alpha } \,c L-L x+x^2)+\sqrt{\alpha } \,c \,L\, e^{\frac{2 x}{\sqrt{\alpha } c}})
\end{split}
\label{v4}
\end{equation}
with
\begin{equation}
\Aq =\, L\, (-\sqrt{\alpha }-(\sqrt{\alpha }+1) \,e^{\frac{L}{\sqrt{\alpha } c}}+1)+2 \,(\alpha -1) \,c \,(e^{\frac{L}{\sqrt{\alpha } c}}-1)
\end{equation}

\break

\section{Inconsistency of limiting solutions}
\label{sec: Inconsistencies}

\def\vL{v_{l}}
\def\chiL{\chi_{l}}
\def\x{\xi}
\def\lc{\lambda}
\def\bv{\bar{\vL}}
\def\bf{\bar{\varphi_l}}
\def\bM{\bar{M_l}}
\def\bT{\bar{T_l}}
\def\bq{\bar{q_l}}

As enlightened by theoretical outcomes in Sect.\ref{sec: LimitingCase}, no solution can be found to Eq.\eqref{MixConv} for a vanishing mixture parameter, i.e.: $\, \m=0 \,$. 
Indeed, for any value $\, c > 0 \,$, constitutive boundary conditions in Eq.\eqref{BCS} can be satisfied by equilibrated bending moments only for a strictly positive mixture parameter.
Hence, inconsistencies of two-phase solutions are expected to occur as $\, \m \to 0 \,$.

In the following, analytical limiting responses of case-studies in Sec.\ref{sec: CaseStudies} are provided. Non-dimensional variables are used in parametric plots by introducing the 
non-dimensional abscissa $\, \xi = x/L \,$, the nonlocal parameter 
$\, \lambda = c/L \,$ and the following non-dimensional limiting displacement fields
\begin{equation}
\bar{\vL} = \vL \frac{K}{F L^3} \quad or \quad \bar{\vL} = \vL \frac{K}{\coppia L^2} \quad or \quad \bar{\vL} = \vL \frac{K}{q L^4}  
\end{equation}

\medskip
\textit{Cantilever under concentrated couple at free end}. 

\begin{figure}[!h]
\centering	
\includegraphics[width=0.8\textwidth]{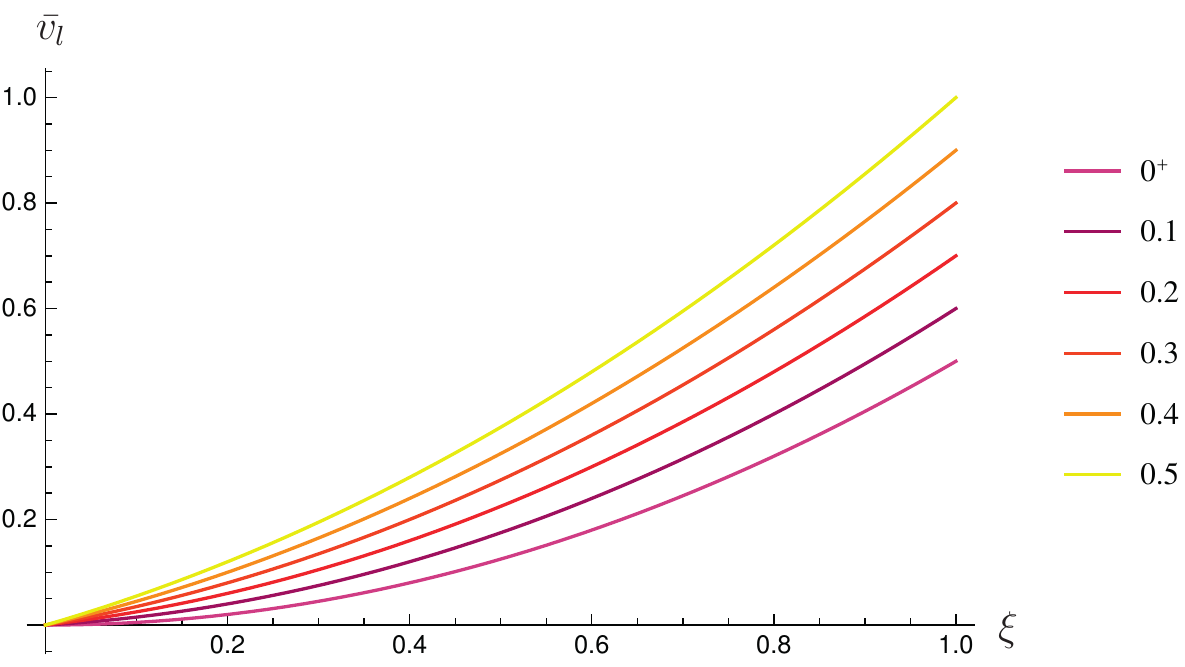}
\caption{Cantilever under concentrated couple at free end: displacement $\,\bv\,$ versus $\,\x\,$ for $\,\lambda \in \{0^+, 0.1, 0.2, 0.3, 0.4, 0.5\}\,$.}
\label{mensolaC_v}
\end{figure}

\begin{figure}[!h]
\centering	
\includegraphics[width=0.8\textwidth]{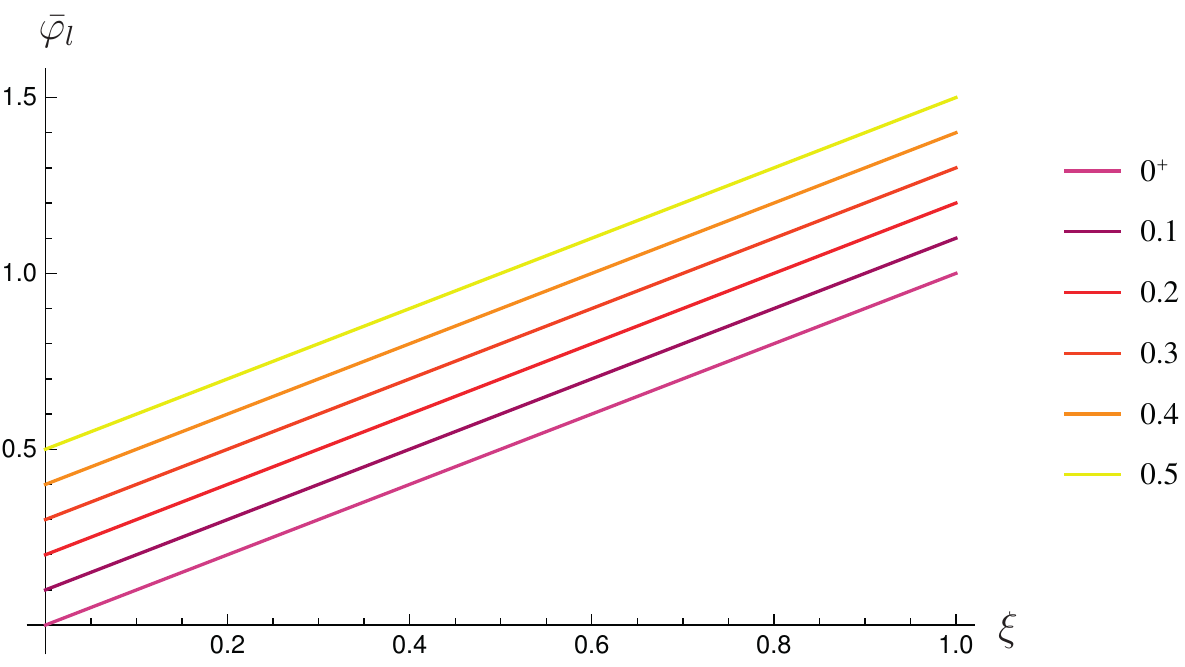}
\caption{Cantilever under concentrated couple at free end: rotation $\,\bf\,$ versus $\,\x\,$ for $\,\lambda \in \{0^+, 0.1, 0.2, 0.3, 0.4, 0.5\}\,$.}
\label{mensolaC_f}
\end{figure}

Computing the limit as $\, \m \to 0 \,$ of Eq.\eqref{v1}, we get  the following 
limiting displacement field
\begin{equation}
\vL(x) = \frac{\coppia}{K}\bigg{(}\frac{x^2}{2} + c\,x\bigg{)}
\label{vLim1}
\end{equation}
The displacement field in Eq.\eqref{vLim1} is clearly in contrast with essential boundary conditions prescribed by constraints. Indeed, except for the local case  $\, \lambda = 0^+ \,$, kinematic inconsistency is apparent in parametric plots of displacement (Fig.\ref{mensolaC_v}) and  rotation fields (Fig.\ref{mensolaC_f}) obtained from derivation of Eq.\eqref{vLim1}.

\begin{figure}[!h]
\centering	
\includegraphics[width=0.8\textwidth]{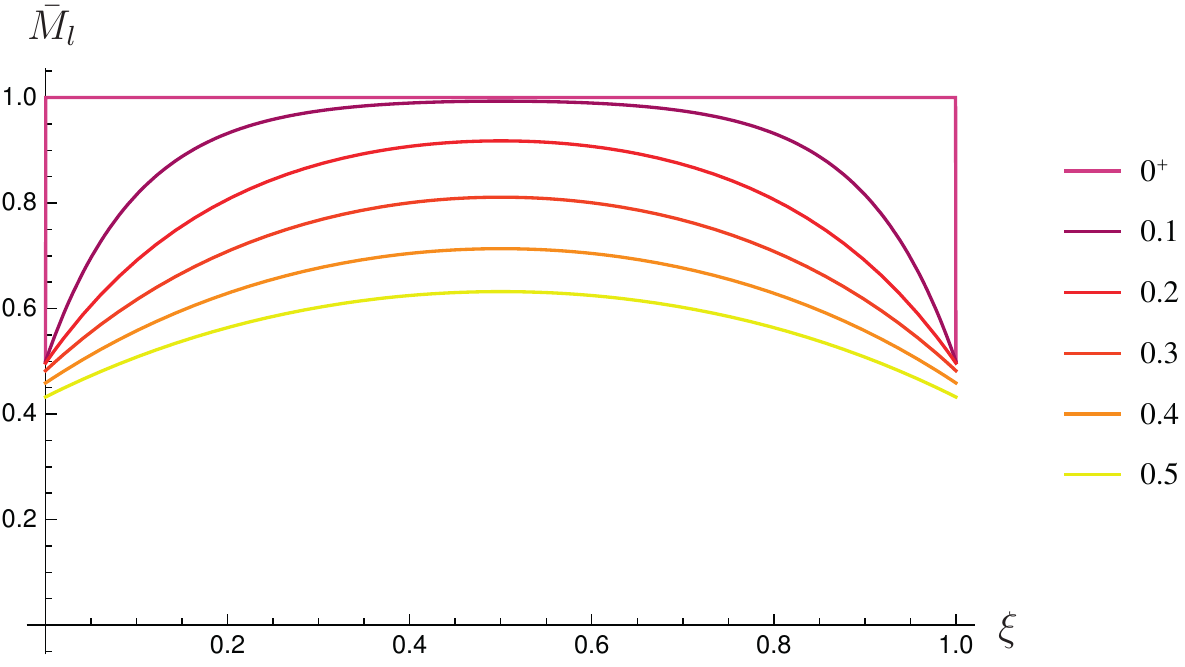}
\caption{Cantilever under concentrated couple at free end: bending moment $\,\bM\,$ versus $\,\x\,$ for $\,\lambda \in \{0^+, 0.1, 0.2, 0.3, 0.4, 0.5\}\,$.}
\label{mensolaC_M}
\end{figure}

\begin{figure}[!h]
\centering	
\includegraphics[width=0.8\textwidth]{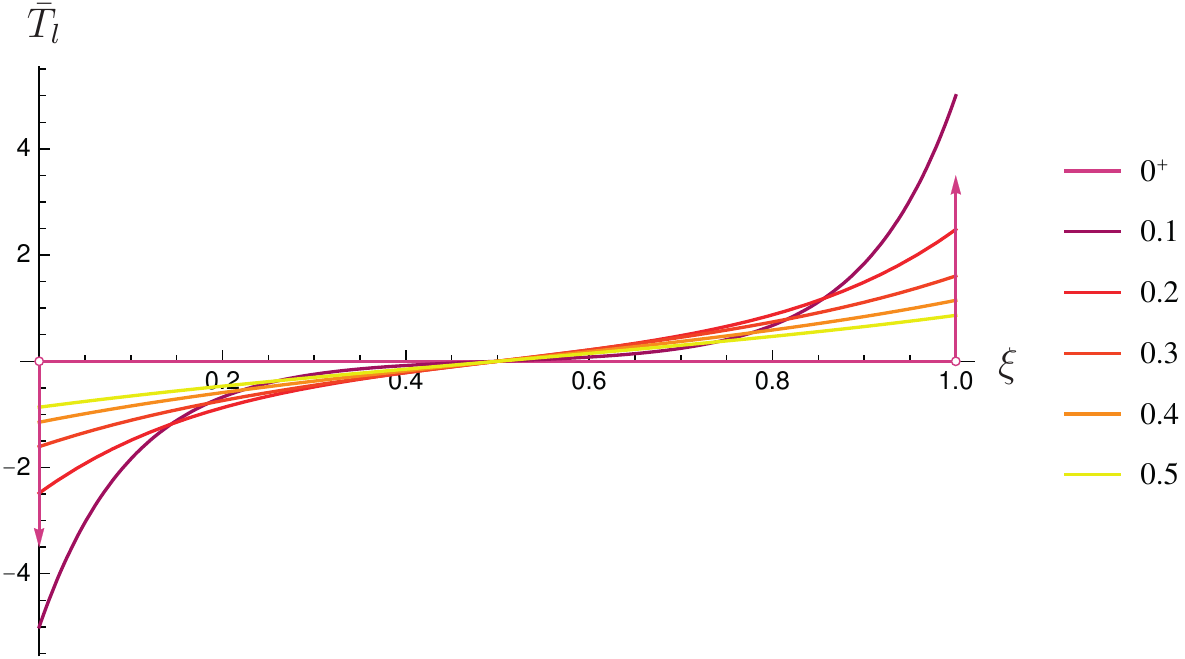}
\caption{Cantilever under concentrated couple at free end: shear force $\,\bT\,$ versus $\,\x\,$ for $\,\lambda \in \{0^+, 0.1, 0.2, 0.3, 0.4, 0.5\}\,$.}
\label{mensolaC_T}
\end{figure}

Double derivation of Eq.\eqref{vLim1} leads to the limiting elastic bending curvature $\, \chiL \,$ that inserted into the integral convolution in Eq.\eqref{MixConv} (with $\, \m = 0 \,$)
should provide the equilibrated bending moment. Moreover, since the redundancy degree is zero, equilibrated bending interaction is the uniform field univocally determined by equilibrium conditions.
As clearly shown by parametric plots in Fig.\ref{mensolaC_M}, the limiting elastic nonlocal curvature $\, \chiL \,$ provides a bending moment which is not equilibrated;
indeed, it is not uniform and not equal to the applied couple, except for the asymptotic local bending moment which is equilibrated for $\, \xi \in \, ]0, 1[ \,$.
A further derivation leads to shear force field which is not vanishing (see Fig.\ref{mensolaC_T}), except for $\, \lambda = 0^+ \,$ with $\, \xi \in \, ]0, 1[ \,$.

\break

\medskip
\textit{Cantilever under concentrated force at free end}. 

\begin{figure}[!h]
\centering	
\includegraphics[width=0.8\textwidth]{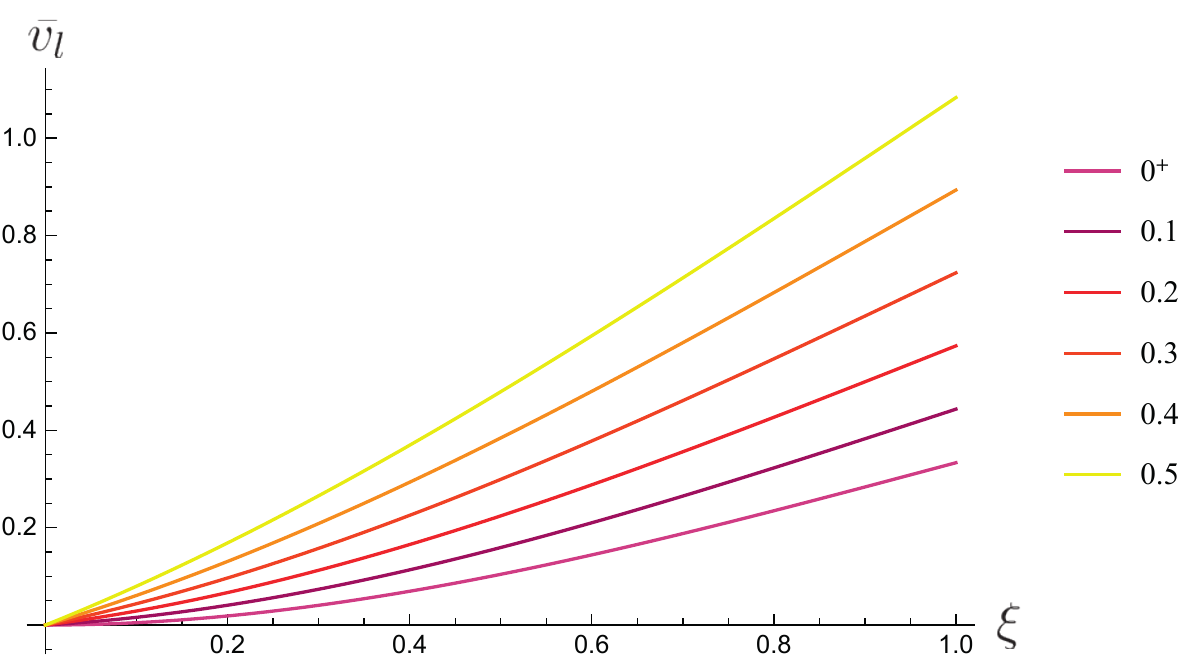}
\caption{Cantilever under concentrated force at free end: displacement $\,\bv\,$ versus $\,\x\,$ for $\,\lambda \in \{0^+, 0.1, 0.2, 0.3, 0.4, 0.5\}\,$.}
\label{mensolaF_v}
\end{figure}

\begin{figure}[!h]
\centering	
\includegraphics[width=0.8\textwidth]{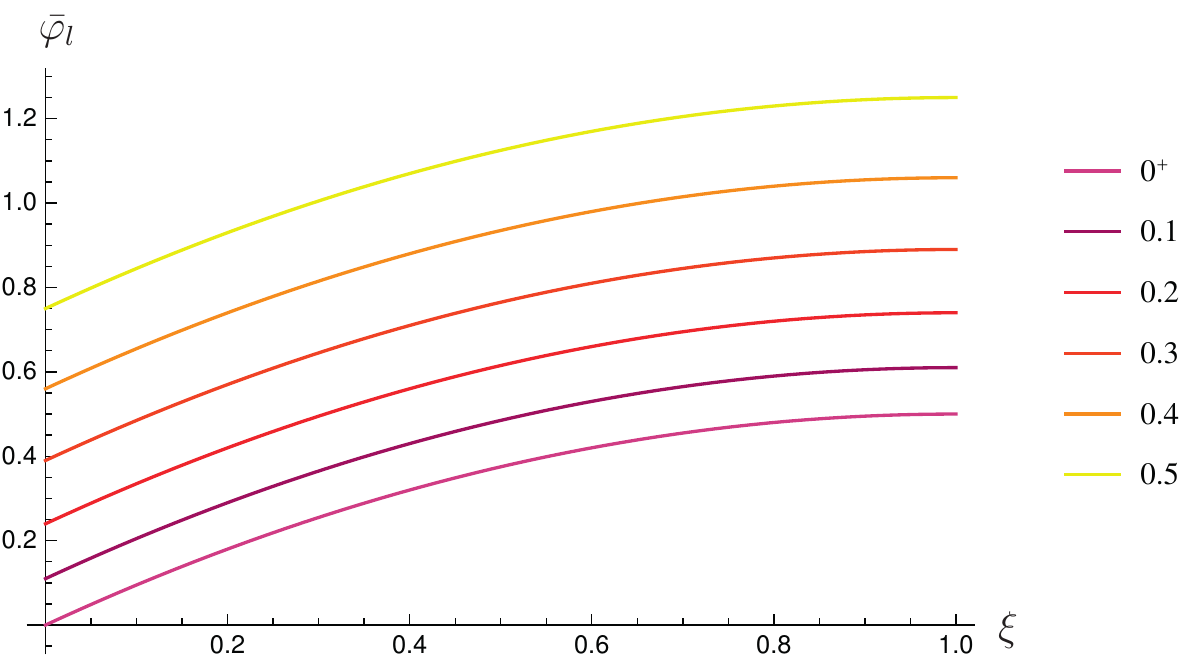}
\caption{Cantilever under concentrated force at free end: rotation $\,\bf\,$ versus $\,\x\,$ for $\,\lambda \in \{0^+, 0.1, 0.2, 0.3, 0.4, 0.5\}\,$.}
\label{mensolaF_f}
\end{figure}

Computing the limit as $\, \m \to 0 \,$ of Eq.\eqref{v2}, we get the following limiting displacement field
%
\begin{equation}
\vL(x) = \frac{F}{K}\bigg{(}L\,\frac{x^2}{2} - \frac{x^3}{6} + c^2\,x + L\,c\,x \bigg{)}
\label{vLim2}
\end{equation}
Kinematic inconsistencies are clearly shown in parametric plots of displacement and rotation fields as a function of $\, \lc \,$, except for the local case $\, \lambda = 0^+ \,$ (see Figs.\ref{mensolaF_v}-\ref{mensolaF_f}).
Double derivation of Eq.\eqref{vLim2} leads to the limiting bending curvature that inserted into the integral convolution in Eq.\eqref{MixConv}, for $\, \m = 0 \,$, provides a bending moment field which is
not compatible with differential and boundary equilibrium requirements.

\begin{figure}[!h]
\centering	
\includegraphics[width=0.8\textwidth]{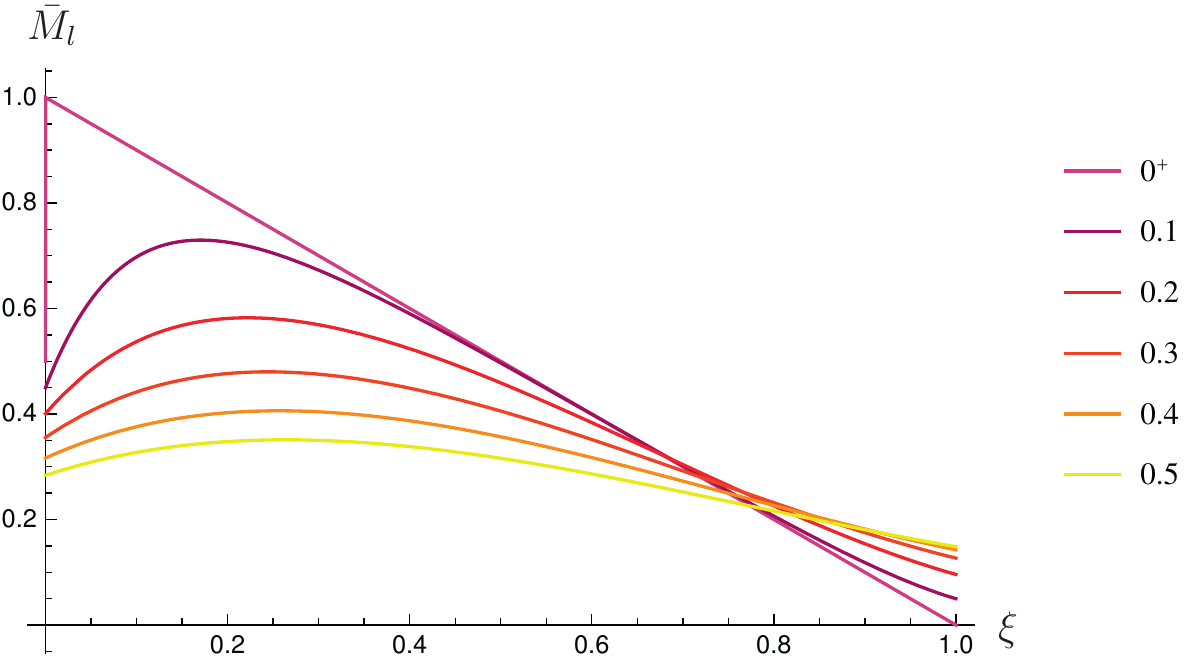}
\caption{Cantilever under concentrated force at free end: bending moment $\,\bM\,$ versus $\,\x\,$ for $\,\lambda \in \{0^+, 0.1, 0.2, 0.3, 0.4, 0.5\}\,$.}
\label{mensolaF_M}
\end{figure}

\begin{figure}[!h]
\centering	
\includegraphics[width=0.8\textwidth]{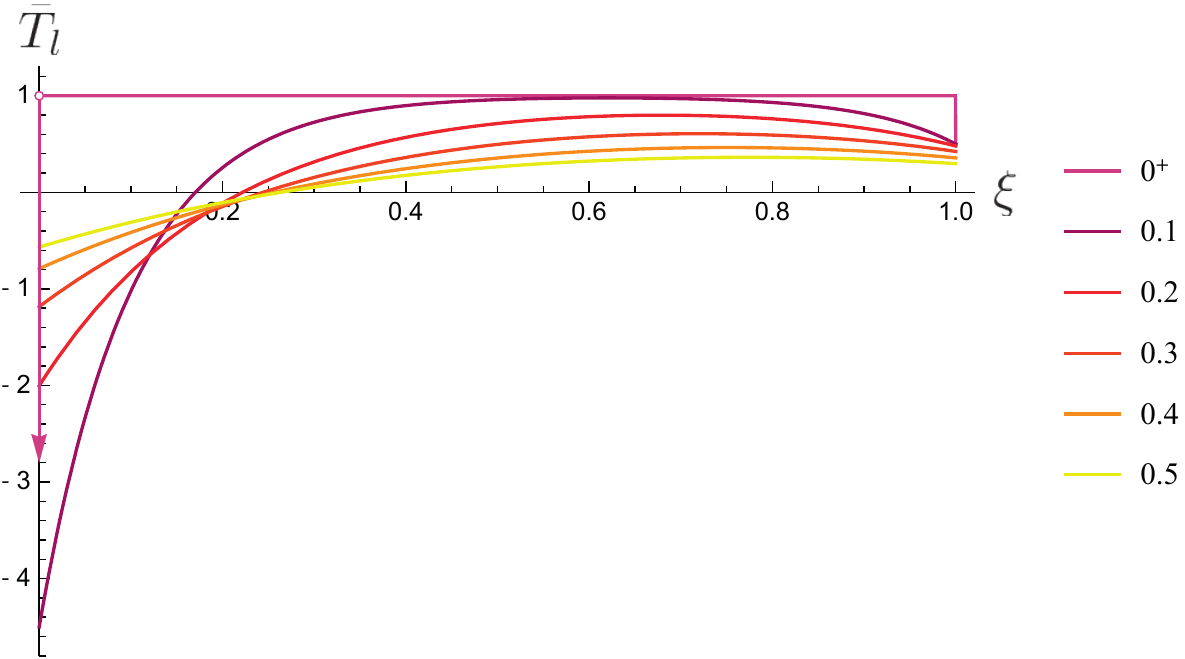}
\caption{Cantilever under concentrated force at free end: shear force $\,\bT\,$ versus $\,\x\,$ for $\,\lambda \in \{0^+, 0.1, 0.2, 0.3, 0.4, 0.5\}\,$.}
\label{mensolaF_T}
\end{figure}

Indeed, since the redundancy degree is zero, the equilibrated bending moment is the linear field univocally determined by equilibrium conditions.
Instead, as clearly shown by parametric plots in Fig.\ref{mensolaF_M}, the limiting elastic bending curvature $\, \chiL \,$ provides a bending moment which is not equilibrated
since it is not linear and does not vanish at free-end of the cantilever, except for the asymptotic local bending moment which is equilibrated for $\, \xi \in \, ]0, 1] \,$.
A further derivation leads to the shear force field which is not equal to the 
value of the applied force (see Fig.\ref{mensolaF_T}) and hence, it is not uniform (except for $\, \lambda = 0^+ \,$ with $\, \xi \in \, ]0, 1[ \,$), i.e.: the emerging distributed loading is not vanishing.

\break

\medskip
\textit{Simply supported beam under uniformly distributed loading}. 

\begin{figure}[!h]
\centering	
\includegraphics[width=0.8\textwidth]{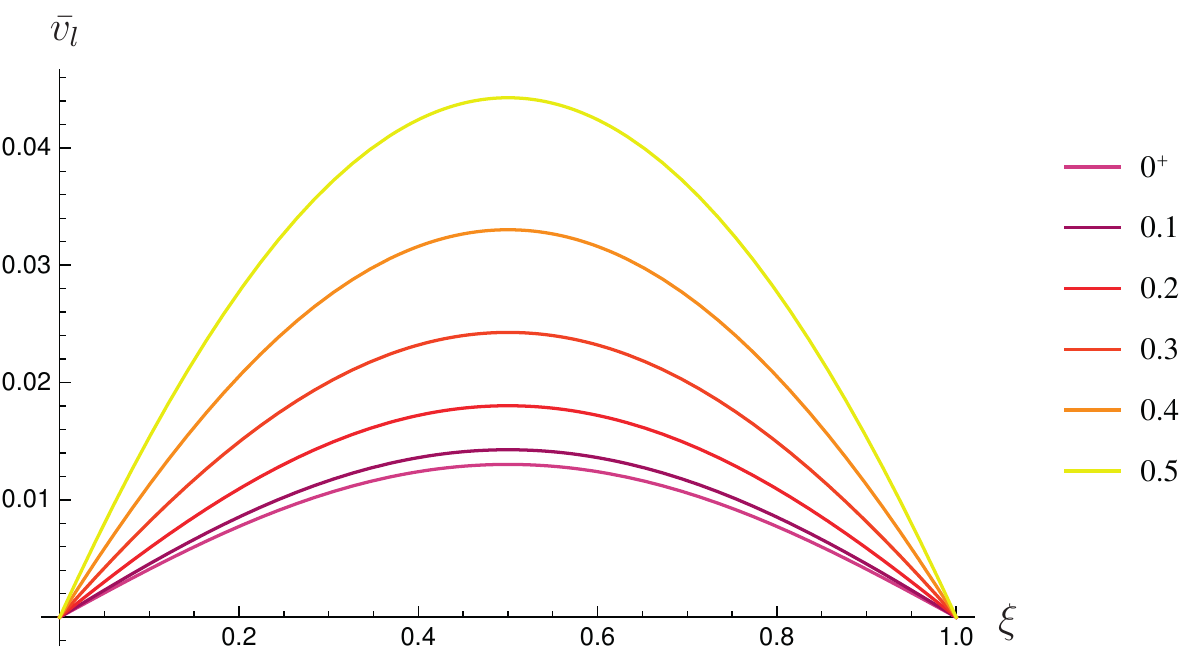}
\caption{Simply supported beam under uniformly distributed loading: displacement $\,\bv\,$ versus $\,\x\,$ for $\,\lambda \in \{0^+, 0.1, 0.2, 0.3, 0.4, 0.5\}\,$.}
\label{isoq_v}
\end{figure}

\begin{figure}[!h]
\centering	
\includegraphics[width=0.8\textwidth]{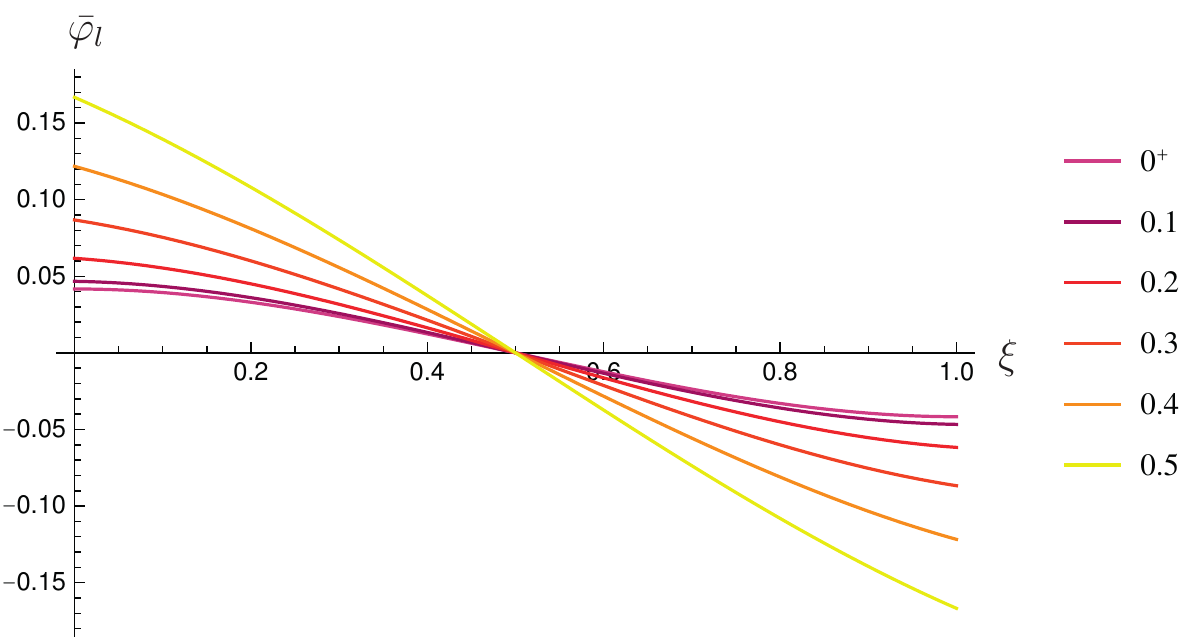}
\caption{Simply supported beam under uniformly distributed loading: rotation $\,\bf\,$ versus $\,\x\,$ for $\,\lambda \in \{0^+, 0.1, 0.2, 0.3, 0.4, 0.5\}\,$.}
\label{isoq_f}
\end{figure}

\begin{figure}[!h]
\centering	
\includegraphics[width=0.8\textwidth]{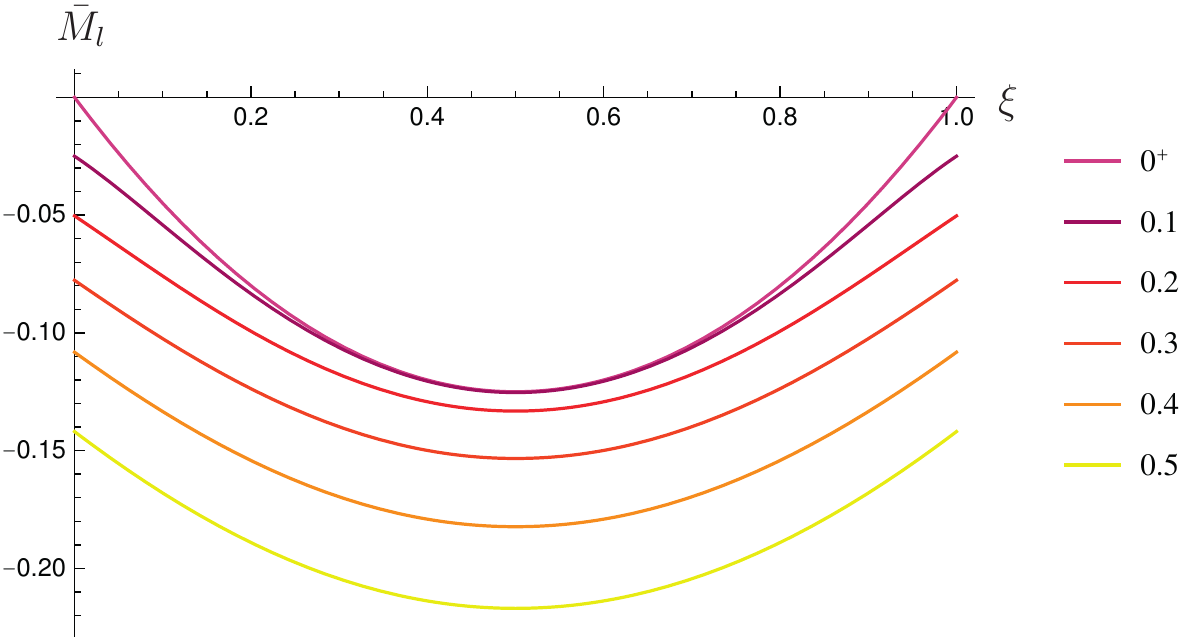}
\caption{Simply supported beam under uniformly distributed loading: bending moment $\,\bM\,$ versus $\,\x\,$ for $\,\lambda \in \{0^+, 0.1, 0.2, 0.3, 0.4, 0.5\}\,$.}
\label{isoq_M}
\end{figure}

\begin{figure}[!h]
\centering	
\includegraphics[width=0.8\textwidth]{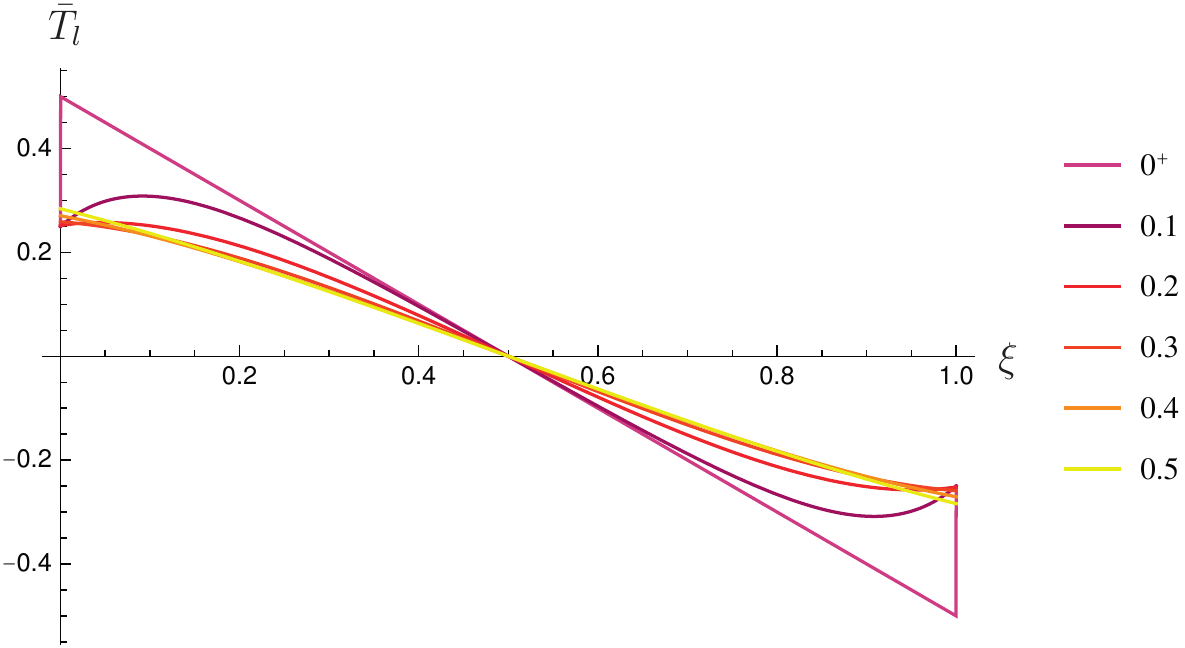}
\caption{Simply supported beam under uniformly distributed loading: shear force $\,\bT\,$ versus $\,\x\,$ for $\,\lambda \in \{0^+, 0.1, 0.2, 0.3, 0.4, 0.5\}\,$.}
\label{isoq_T}
\end{figure}

\begin{figure}[!h]
\centering	
\includegraphics[width=0.8\textwidth]{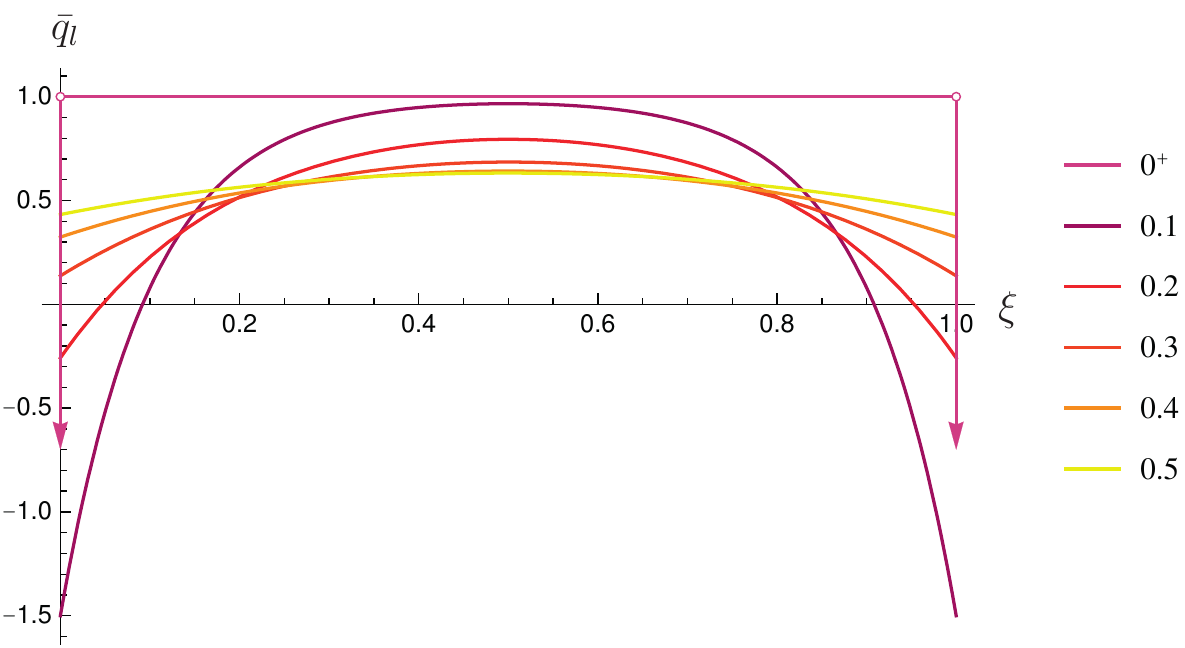}
\caption{Simply supported beam under uniformly distributed loading: emerging loading $\,\bq\,$ versus $\,\x\,$ for $\,\lambda \in \{0^+, 0.1, 0.2, 0.3, 0.4, 0.5\}\,$.}
\label{isoq_q}
\end{figure}

Computing the limit as $\, \m \to 0 \,$ of Eq.\eqref{v3}, we get the following limiting displacement 
\begin{equation}
\vL(x) = \frac{q}{K} \bigg{(}\frac{x^4}{24} -\frac{L x^3}{12} -\frac{c^2 x^2}{2}+\frac{L^3 x}{24} (12 \lambda ^2+1)\bigg{)}
\label{vLim3}
\end{equation}
Parametric plots of displacement and rotation fields as function of $\, \lc \,$ are shown in Figs.\ref{isoq_v}-\ref{isoq_f}. Double derivation of Eq.\eqref{vLim3} leads to the limiting bending curvature that inserted into the integral convolution in Eq.\eqref{MixConv} (for $\, \m = 0 \,$) provides the static fields shown in Figs.\ref{isoq_M}-\ref{isoq_T}. 
The bending moment field is clearly incompatible with natural boundary conditions and the shear force field is not linear, i.e.: the emerging distributed loading is not uniform (see Fig.\ref{isoq_q}). Hence, the interaction fields for $\, \lc > 0\,$ are not compatible with differential and boundary equilibrium requirements.

\break

\medskip
\textit{Doubly clamped beam under uniformly distributed loading}. 

The limiting displacement field as $\, \m \to 0 \,$ of Eq.\eqref{v4} is given by
\begin{equation}
\vL(x) = \frac{q}{K}\left(\frac{x^4}{24}-\frac{L x^3}{12}-\frac{c^2 x^2}{2}+\frac{c \left(12 \lambda ^2+6 \lambda +1\right) L^2 x}{12 (2 \lambda +1)}+\frac{L^2 x^2}{24 (2 \lambda +1)}\right)
\label{vLim4}
\end{equation}
\break

\begin{figure}[!h]
\centering	
\includegraphics[width=0.8\textwidth]{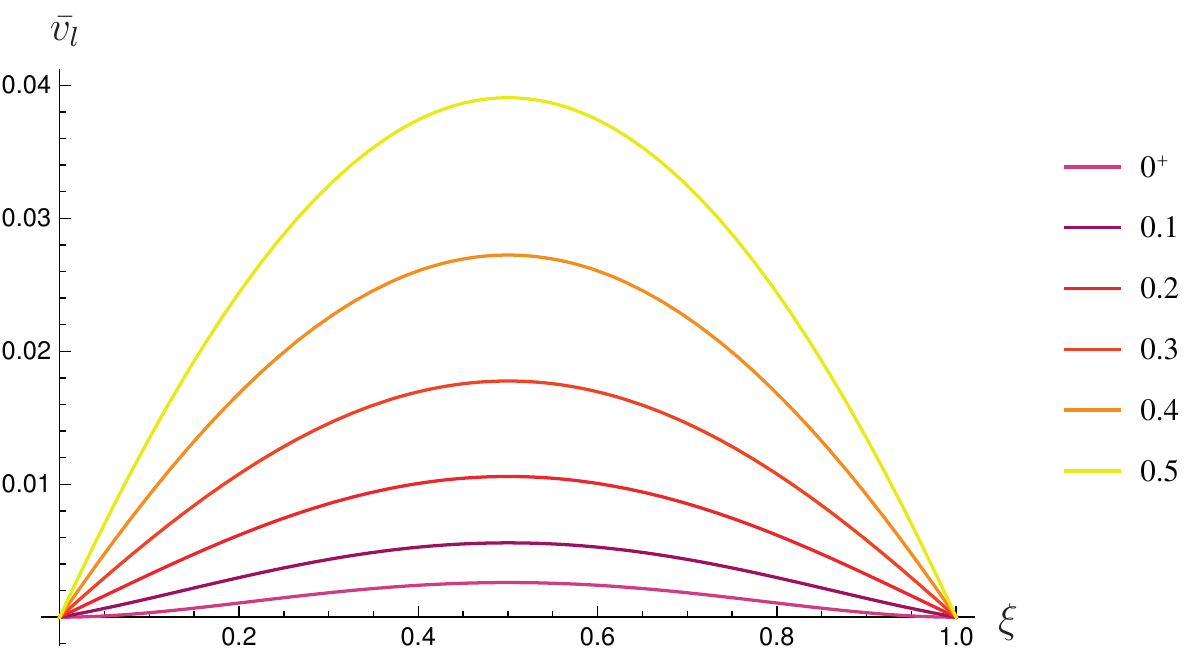}
\caption{Doubly clamped beam under uniformly distributed loading: displacement $\,\bv\,$ versus $\,\x\,$ for $\,\lambda \in \{0^+, 0.1, 0.2, 0.3, 0.4, 0.5\}\,$.}
\label{ipeq_v}
\end{figure}

\begin{figure}[!h]
\centering	
\includegraphics[width=0.8\textwidth]{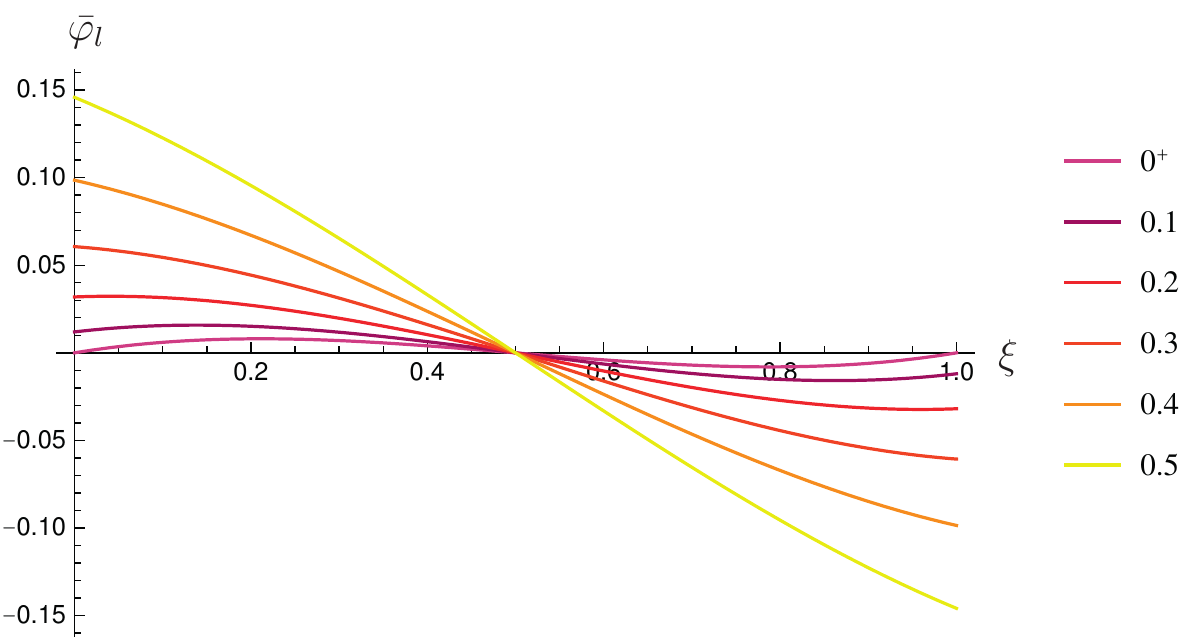}
\caption{Doubly clamped beam under uniformly distributed loading: rotation $\,\bf\,$ versus $\,\x\,$ for $\,\lambda \in \{0^+, 0.1, 0.2, 0.3, 0.4, 0.5\}\,$.}
\label{ipeq_f}
\end{figure}

\begin{figure}[!h]
\centering	
\includegraphics[width=0.8\textwidth]{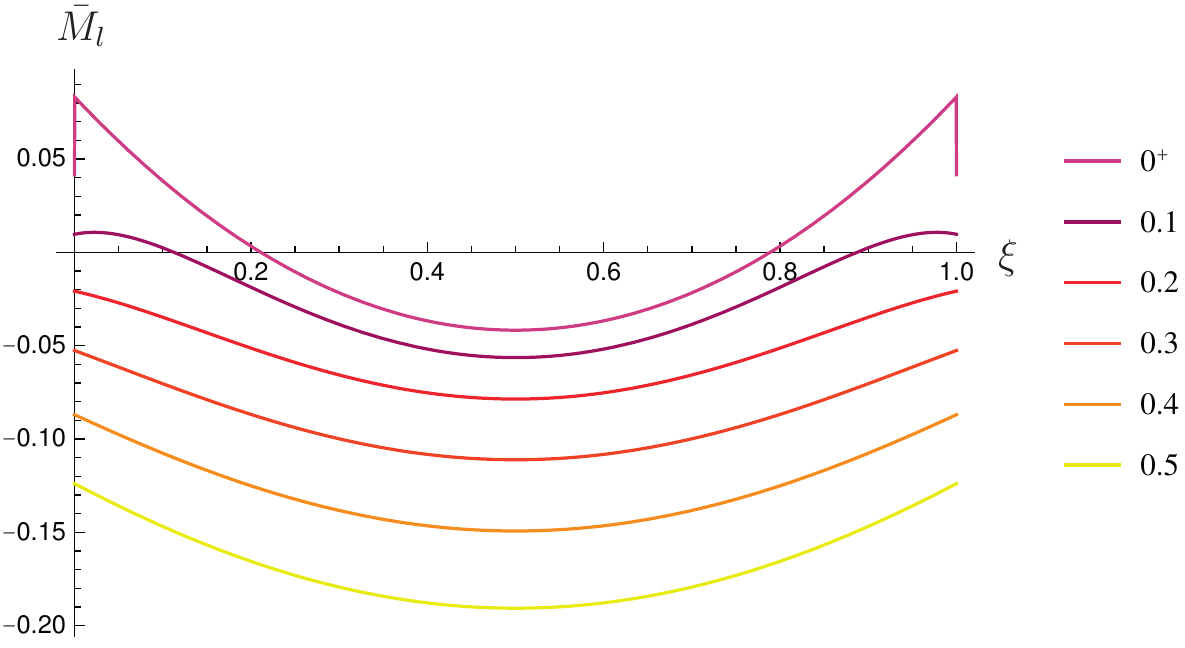}
\caption{Doubly clamped beam under uniformly distributed loading: bending moment $\,\bM\,$ versus $\,\x\,$ for $\,\lambda \in \{0^+, 0.1, 0.2, 0.3, 0.4, 0.5\}\,$.}
\label{ipeq_M}
\end{figure}

\begin{figure}[!h]
\centering	
\includegraphics[width=0.8\textwidth]{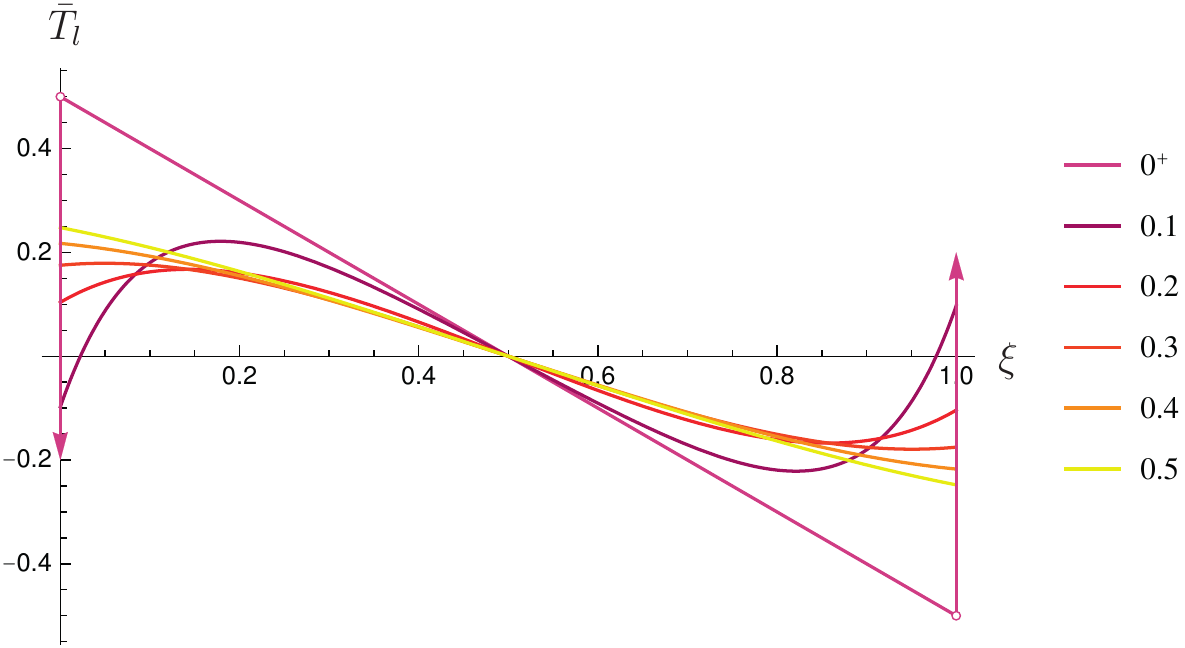}
\caption{Doubly clamped beam under uniformly distributed loading: shear force $\,\bT\,$ versus $\,\x\,$ for $\,\lambda \in \{0^+, 0.1, 0.2, 0.3, 0.4, 0.5\}\,$.}
\label{ipeq_T}
\end{figure}

\begin{figure}[!h]
\centering	
\includegraphics[width=0.8\textwidth]{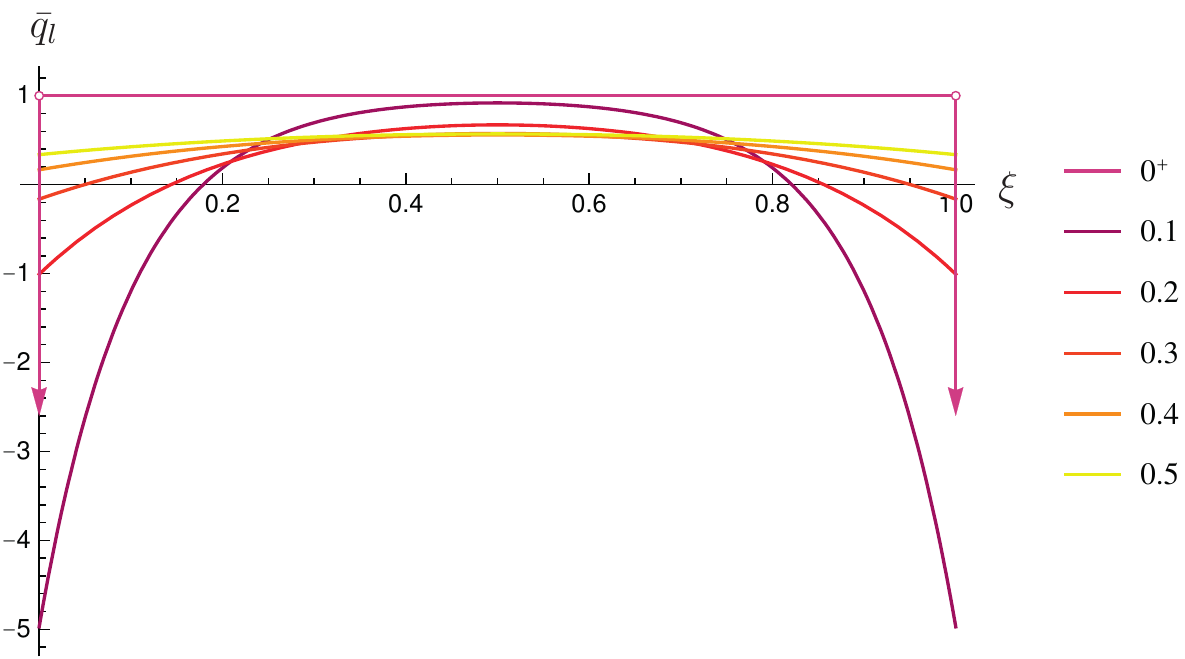}
\caption{Doubly clamped beam under uniformly distributed loading: emerging loading $\,\bq\,$ versus $\,\x\,$ for $\,\lambda \in \{0^+, 0.1, 0.2, 0.3, 0.4, 0.5\}\,$.}
\label{ipeq_q}
\end{figure}

The displacement field in Eq.\eqref{vLim4} is in contrast with essential boundary conditions prescribed by constraints. These kinematic inconsistencies are clearly shown in parametric plots of displacement and rotation fields (see Figs.\ref{ipeq_v}-\ref{ipeq_f}). 
Double derivation of Eq.\eqref{vLim2} leads to the limiting nonlocal elastic curvature that is used to compute the integral convolution in Eq.\eqref{MixConv}, for $\, \m = 0 \,$, to get the bending moment field depicted in Fig.\ref{ipeq_M}.
A further derivation leads to shear force field which is not linear (see Fig.\ref{ipeq_T}) and the emerging distributed loading is not uniform and not equal to the applied one (see Fig.\ref{ipeq_q}).
Hence, limiting solutions of the elastostatic problem do not satisfy kinematic compatibility and equilibrium requirements, except for the asymptotic local displacement and rotation fields $\,\forall \xi \in \, [0, 1] \,$ and for
the asymptotic local bending and shearing fields for $\,\xi \in \, ]0, 1[ \,$.  

\section{Closing remarks}
\label{sec: Concl}

The bending behaviour of elastic beams has been investigated by Eringen's two-phase integral theory.
Exact closed-form solutions of simple structural problems of applicative interest have been preliminarily provided in terms of mixture and nonlocal parameters.
Then, relevant asymptotic fields (as the mixture parameter tends to zero) have been evaluated and studied. 
It has been proven that such fields are in contrast with kinematic boundary conditions and
equilibrium requirements for any nonlocal parameter value. 
Accordingly, limiting responses of Eringen's two-phase formulation cannot be used as
solutions of the ill-posed Eringen purely nonlocal structural problem.
Therefore, inaccurate outcomes contributed in current literature \citep[see e.g.][]{Mikhasev2020,Mikhasev2021}, regarding 
the well-posedness of limiting elastodynamic problems 
based on Eringen's two-phase continuum theory, should be amended consequently.

\medskip
\noindent
\textbf{Acknowledgment} -
Financial support from the MIUR in the framework of the Project PRIN 2017 - code 2017J4EAYB \emph{Multiscale Innovative Materials and Structures (MIMS)}; University of Naples Federico II Research Unit - is gratefully acknowledged.

\break


\begin{thebibliography}{00}




\bibitem[Banejad et al.(2020)]{Banejad2020}
Banejad, A., Passandideh-Fard, M., Niknam, H., {Mirshojaeian Hosseini}, M.J.,
  {Mousavi Shaegh}, S.A., 2020. Design, fabrication and experimental characterization
  of whole-thermoplastic microvalves and micropumps having micromilled liquid
  channels of rectangular and half-elliptical cross-sections.
Sensors and Actuators A 301, 111713.
\bibitem[Barretta et al.(2018a)]{BarrettaPhysicaE}
Barretta, R., Fabbrocino, F., Luciano, R., Marotti de Sciarra, F., 2018a. Closed-form solutions in stress-driven two-phase integral elasticity for bending of functionally graded nano-beams. Physica E: Low-dimensional Systems and Nanostructures 97, 13–30.
\bibitem[Barretta et al.(2018b)]{BarrettaCanJCOMB2018}
Barretta, R., \v{C}anadija, M., Feo, L., Luciano, R., Marotti de Sciarra, F., Penna, R., 2018b.
Exact solutions of inflected functionally graded nano-beams in integral elasticity. 
Composites Part B 142, 273-286.
\bibitem[Barretta et al.(2018c)]{BarrettaMRC2018}
Barretta, R., Diaco, M., Feo, L., Luciano, R., Marotti de Sciarra, F., Penna, R., 2018c.
Stress-driven integral elastic theory for torsion of nano-beams. 
Mechanics Research Communications 87, 35-41.
\bibitem[Barretta et al.(2019)]{Barretta2019}
Barretta, R., Marotti de Sciarra, F., Vaccaro, M.S., 2019. 
On nonlocal mechanics of curved elastic beams. 
International Journal of Engineering Science 144, 103140.
\bibitem[Ba\v{z}ant and Jir\'asek(2002)]{Bazant2002}
Ba\v{z}ant, Z.P., Jir\'asek, M., 2002.
Nonlocal integral formulation of plasticity and damage: survey of progress. 
Journal of Engineering Mechanics - ASCE, 128, 1119-1149.
\bibitem[Borino et al.(2003)]{Borino2003}
Borino, G., Failla, B., Parrinello, F., 2003. 
A symmetric nonlocal damage theory. 
International Journal of Solids and Structures 40, 3621-3645.
\bibitem[Chao et al.(2020)]{Chao2020}
Chao, M., Wang, Y., Ma, D., Wu, X., Zhang, W., Zhang, L., Wan, P., 2020. 
Wearable mxene nanocomposites-based strain sensor with tile-like stacked hierarchical
microstructure for broad-range ultrasensitive sensing.
Nano Energy 78, 105187.
\bibitem[Chorsi and Chorsi(2018)]{Chorsi2018}
Chorsi, M.T. , Chorsi, H.T., 2018. 
Modeling and analysis of MEMS disk resonators. 
Microsystem Technologies 24 (6), 2517–2528 .
\bibitem[Eringen(1972)]{Eringen1972}
Eringen, A. C., 1972.
Linear theory of nonlocal elasticity and dispersion of plane waves.
International Journal of Engineering Science 10(5), 425-435. 
\bibitem[Eringen(1983)]{Eringen1983}
Eringen, A. C., 1983.
On differential equations of nonlocal elasticity and 
solutions of screw dislocation and surface waves.
Journal of Applied Physics 54, 4703.
\bibitem[Eringen(1987)]{Eringen1987}
Eringen, A. C., 1987.
Theory of nonlocal elasticity and some applications. 
Res Mechanica 21, 313-342.
\bibitem[Eptaimeros et al.(2016)]{Eptaimeros2016}
Eptaimeros, K.G., Koutsoumaris, C.C., Tsamasphyros, G.J., 2016. Nonlocal integral approach to the dynamical response of nano-beams. Int J Mech Sci 115–116, 68–80 .
\bibitem[Fathi and Ghassemi(2017)]{Fathi2017}
Fathi, M., Ghassemi, A. (2017). 
The effects of surface stress and nonlocal small scale on the uniaxial and biaxial buckling of the rectangular piezoelectric nanoplate based on the two variable-refined plate theory.
Journal of the Brazilian Society of Mechanical Sciences and Engineering 39 3203-3216.
\bibitem[Fern{\'a}ndez-S{\'a}ez and Zaera(2017)]{FernandezZaera2017}
Fern{\'a}ndez-S{\'a}ez, J., Zaera, R., 2017.
Vibrations of Bernoulli-Euler beams using the two-phase nonlocal elasticity theory. 
International Journal of Engineering Science 119, 232-248.
\bibitem[Ghayesh and Farajpour(2019)]{GhayeshIJESciS2019}
Ghayesh, M.H., Farajpour, A., 2019.
A review on the mechanics of functionally graded nanoscale and microscale structures.
International Journal of Engineering Science 137, 8-36.
\bibitem[Ghayesh and Farokhi(2020)]{Ghayesh2020}
Ghayesh, M.H., Farokhi, H., 2020.
Nonlinear broadband performance
of energy harvesters. Int J Eng Sci 147, 103202
\bibitem[Jankowski et al.(2020)]{Jankowski2020}
Jankowski, P., \.{Z}ur, K.K., Kim, J., Reddy, J.N., 2020.
On the bifurcation buckling and vibration of porous nanobeams. Composite Structures 250, 112632.
\bibitem[Karami et al.(2018)]{Karami2018}
Karami, B., Shahsavari, D., Janghorban, M., Li, L., 2018.
Wave dispersion of mounted graphene with initial stress.
Thin-Walled Structures 122, 102-111.
\bibitem[Kiani(2021)]{Kiani2021}
Kiani, K., \.{Z}ur, K.K., 2021.
Vibrations of double-nanorod-systems with defects using nonlocal-integralsurface
energy-based formulations.
Composite Structures 256, 113028.
\bibitem[Khodabakhshi and Reddy(2015)]{Khodabakhshi2015}
Khodabakhshi, P., Reddy, J.N., 2015. A unified integro-differential nonlocal model.
International Journal of Engineering Science 95, 60-75.
\bibitem[Kr\"oner et al.(1967)]{Kroner1967}
Kr\"oner, E., 1967.
Elasticity theory of materials with long range cohesive forces. 
International Journal of Solids and Structures 3(5), 731-742.
\bibitem[Lu et al.(2019)]{LuP2019}
Lu, P., Veletić, M., Laasmaa, M., Vendelin, M., Louch, W.E., Steinar Halvorsen P., Bergsland, J., Balasingham, I., 2019. Multi-nodal nano-actuator pacemaker for energy-efficient stimulation of cardiomyocytes. Nano Communication Networks 22, 100270.
\bibitem[Malikan et al.(2020b)]{Malikan2020b}
Malikan, M., Eremeyev, V.A., \.{Z}ur, K.K., 2020.
Effect of axial porosities on flexomagnetic response
of in-plane compressed piezomagnetic nanobeams.
Symmetry 12(12), 1935.
\bibitem[Mikhasev and Nobili(2020)]{Mikhasev2020}
Mikhasev, G., Nobili, A., 2020. On the solution of the purely nonlocal theory of beam elasticity as a limiting case of the two-phase theory. International Journal of Solids and Structures 190, 47–57.
\bibitem[Mikhasev(2021)]{Mikhasev2021}
Mikhasev, G., 2021.
Free high-frequency vibrations of nonlocally elastic beam with varying cross-section area.
Continuum Mechanics and Thermodynamics. 
https://doi.org/10.1007/s00161-021-00977-6
\bibitem[Oskouie et al.(2018)]{AnsariActaMechSin2018}
Oskouie, M. F., Ansari, R., Rouhi, H., 2018.
Bending of Euler-Bernoulli nanobeams based on the strain-driven 
and stress-driven nonlocal integral models: a numerical approach.
Acta Mechanica Sinica.
Doi: 10.1007/s10409-018-0757-0
\bibitem[Pinnola(2020)]{Pinnola2020}
Pinnola, F.P., Vaccaro, M.S., Barretta, R., Marotti de Sciarra, F., 2020. Random vibrations
of stress-driven nonlocal beams with external damping. Meccanica URL https:
//doi.org/10.1007/s11012-020-01181-7.
\bibitem[Pisano and Fuschi(2003)]{PisanoFuschi2003}
Pisano, A.A., Fuschi, P., 2003.
Closed form solution for a nonlocal elastic bar in tension.
International Journal of Solids and Structures 40, 13-23. 
\bibitem[Pisano et al.(2021)]{PisanoZamm2021}
Pisano, A.A., Fuschi, P., Polizzotto, C., 2021.
Integral and differential approaches to Eringen's nonlocal elasticity models accounting for boundary effects with applications to beams in bending.
ZAMM - Journal of Applied Mathematics and Mechanics. 
https://doi.org/10.1002/zamm.202000152
\bibitem[Polizzotto(2001)]{Polizzotto2001}
Polizzotto, C., 2001. 
Nonlocal elasticity and related variational principles. 
International Journal of Solids and Structures 38, 7359-7380.
\bibitem[Polizzotto(2002)]{Polizzotto2002}
Polizzotto, C., 2002. 
Thermodynamics and continuum fracture mechanics for nonlocal-elastic plastic materials. 
European Journal of Mechanics A/Solids 21, 85-103.
\bibitem[Polyanin and Manzhirov(2008)]{Polyanin2008}
Polyanin, A.D., Manzhirov A.V., 2008.  
Handbook of integral equations. 2nd ed. Boca Raton,
FL: Chapman \& Hall/CRC.
\bibitem[Roghani(2020)]{Roghani2020}
Roghani, M., Rouhi, H., 2020. Nonlinear stress-driven nonlocal formulation of timoshenko
beams made of fgms. Continuum Mechanics and Thermodynamics.  https://doi.org/10.
1007/s00161-020-00906-z
\bibitem[Rogula(1965)]{Rogula1965}
Rogula, D., 1965.
Influence of spatial acoustic dispersion on dynamical properties of dislocations. 
Bulletin de l'Académie Polonaise des Sciences, Séries des Sciences Techniques 13, 337-343 .
\bibitem[Rogula(1982)]{Rogula1982}
Rogula, D., 1982.
Introduction to nonlocal theory of material media.
In D. Rogula (Ed.), \textit{Nonlocal theory of material media}, CISM courses and lectures (vol. 268, pp. 125-222). Wien: Springer.
\bibitem[Romano et al.(2017a)]{Romano2017} 
Romano, G., Barretta, R., Diaco, M., Marotti de Sciarra, F., 2017a.
Constitutive boundary conditions and paradoxes in nonlocal elastic
nano-beams. 
International Journal of Mechanical Sciences 121, 151-156.
\bibitem[Romano et al.(2017b)]{RomanoIntModels2017}
Romano, G., Barretta, R., Diaco, M., 2017b.
On nonlocal integral models for elastic nano-beams. 
International Journal of Mechanical Sciences 131-132, 490-499.
\bibitem[Sedighi et al.(2017)]{Sedighi2020}
Sedighi, H.M., Malikan, M., Valipour, A., \.{Z}ur, K.K., 2020.
Nonlocal vibration of carbon/boron-nitride
nano-hetero-structure in thermal and magnetic fields
by means of nonlinear finite element method.
Journal of Computational Design and Engineering 7(5), 591–602.
\bibitem[Soukarié et al.(2020)]{Soukarie2020}
Soukarié, D., Ecochard, V., Salomé, L., 2020. Dna-based nanobiosensors for
  monitoring of water quality.
Int. J. Hyg. Environ. Health 226, 113485.
\bibitem[Tricomi(1957)]{Tricomi1985}
Tricomi, F.G., 1957. 
Integral Equations. Interscience, New-York, USA.
Reprinted by Dover Books on Mathematics, 1985.
\bibitem[Vila et al.(2017)]{Vila2017}
Vila, J., Fern{\'a}ndez-S{\'a}ez, J., Zaera, R., 2017.
Nonlinear continuum models for the dynamic behavior of 1D microstructured solids.
International Journal of Solids and Structures 117, 111-122.
\bibitem[Wang et al.(2016)]{Wang2016}
Wang, Y., Zhu,  X., Dai, H., 2016.
Exact solutions for the static bending of Euler-Bernoulli beams using 
Eringen two-phase local/nonlocal model.
AIP Advances 6(8), 085114. 
Doi:10.1063/1.4961695
\bibitem[Zhang(2017)]{Zhang2017}
Zhang, Y., 2017.
Frequency spectra of nonlocal Timoshenko beams and an effective method of determining nonlocal effect.
International Journal of Mechanical Sciences 128-129, 572-582.
\bibitem[Zhang(2020)]{Zhang2020}
Zhang, J.Q., Qing, H., Gao, C.F., 2020. Exact and asymptotic bending analysis of microbeams
under different boundary conditions using stress-derived nonlocal integral model. Z.
Angew. Math. Mech. 100(1), e201900148.
\bibitem[Zhu and Li(2017)]{ZhuLi2017}
Zhu, X., Li, L., 2017.
Closed form solution for a nonlocal strain gradient rod in tension.
International Journal of Engineering Science 119, 16-28.
\bibitem[\.{Z}ur et. al(2020)]{Zur2020}
\.{Z}ur, K.K., Arefi, M., Kimc, J., Reddy, J.N., 2020.
Free vibration and buckling analyses of magneto-electro-elastic FGM
nanoplates based on nonlocal modified higher-order sinusoidal shear
deformation theory. Composites Part B 182, 107601.


\end{thebibliography}
\end{document}